\documentstyle[preprint,prd,aps,12pt]{revtex}
\preprint{
$
\begin{array}{r}
\text{hep-ph 9311355} \\ \text{EP-CPTh.A274.1293} \\ \text{LAVAL-PHY-13-93}
\end{array}
$
}
\tighten

\begin{document}
\author{Luc Marleau\cite{Laval}}
\address{Centre de Physique Th\'eorique,\'Ecole
Polytechnique\\91128\\Palaiseau\\CEDEX,\\France}
\title{New classes of summable Skyrmions }
\date{October 1993 }
\maketitle
\draft

\begin{abstract}
We introduce two new classes of summable Skyrmions. The Lagrangians they
originate from are explicitely constructed. We also analyse how some models
could be solve. Exact solutions are found for Skyrme-like toy models.
\end{abstract}

\section{Introduction}

The Lagrangian originally proposed by Skyrme\cite{Skyrme61} is composed of a
kinetic term and a self-interaction term of order two and four,
respectively, in the derivatives of the pion field. The second term prevents
the model from being renormalizable, but it induces pion field
configurations which are soliton solutions (Skyrmions). These are
interpreted as baryons and a number of their properties can be computed
leading to many successfull predictions. The $1/N_c$ analysis however
suggests that if an effective Lagrangian is to represent accurately the
low-energy limit of QCD, it should behave as an effective theory of
infinitely many mesons and include all orders of derivative of the fields
\cite{tHooft74}. In that sense, the Skyrme model represent a naive
(order-four) attempt to provide such a description. The exact form of such
the low-energy effective Lagrangian is for the moment out our reach and
indeed, would be equivalent to a finding a solution for the low-energy limit
of QCD and perhaps to confinement. In the absence of such solutions, one
must rely on more elaborate effective Lagrangians but in doing so one faces
problems: (i) the number of possible extensions, i.e. the number of possible
terms at higher orders becomes increasingly large, (ii) the degree of the
equation of motion becomes infinite and, (iii) adding terms of all orders
introduces an arbitrarily large number of parameters and one loses any
predictive power. Any one of these problems causes the general approach to
become too complex and its treatment intractable.

There is however an alternative. One can impose some kind of symmetry or
constraint that reduces this degree of arbitrariness and complexity but
still retain some interesting new features. The idea is not new. Indeed, the
second term in Skyrme Lagrangian is not the most general one but Skyrme
justified this choice by arguing that this is the only combination which is
of degree two, at most, in the derivative with respect to time. This term
however stabilize the soliton and although it is of order four, it leads to
a nonlinear differential equation of degree two for the static solution
having a hedgehog form. Obviously, the same constraint cannot be implemented
to orders larger then eight if one wants to preserve Lorentz covariance.
Recently, we proposed a different approach to construct tractable all-orders
effective Lagrangian \cite{Marleau89-10}: Using the hedgehog ansatz for
static solution, we required that the degree of the differential equation
remains two. This defines a class of all-orders Skyrmions with a number of
interesting properties: (i) they are more conform to reality since they come
from all-orders effective Lagrangians, (ii) the Lagrangians are chirally
invariant by construction which will eventually allow more control over
symmetry breaking terms since they can be added by hand afterwards, (iii)
the physical predictions can be significantly improved \cite{Marleau90-13},
(iv) their topological and stability properties are similar and (v) they
have been shown to have new interesting features such as a two-phase
structure \cite{Riska93}.

In this work, we identify two more class of tractable all-orders Skyrmions.
This is done in section II, where we also find the explicit form of the
Lagrangians from which they originate. The next section looks at toy models
that can be solved analytically. Although their physical interpretation is
not obvious some of them carry interesting properties like scale invariance
of the soliton. Finally, the last section contains a brief discussion.

\section{All-orders Skyrmions}

\subsection{A class of all-orders Skyrmions}

The Skyrme model is based on the chirally invariant Lagrangian
\begin{equation}
\label{LSkyrme}{\cal L}_{{\rm Skyrme}}=-\frac{f_\pi ^2}4%
\mathop{\rm Tr}
L_\mu L^\mu +\frac{\epsilon _{}^2}4%
\mathop{\rm Tr}
[L_\mu ,L_\nu ]^2
\end{equation}
where $L_\mu =U^{\dagger }\partial _\mu U$ is the left-handed current and $U$
is an SU(2) matrix related to the pion fields by the $U\equiv (\sigma +i{\bf %
\tau \cdot \pi })/f_\pi $. Two parameters are required, $f_\pi $, the pion
decay constant (93 MeV), and $\epsilon $ a dimensionless constant. The first
and second term of this expression is the nonlinear $\sigma $-model
Lagrangian and a stabilizing term added by Skyrme respectively.

The class of effective Lagrangians that we are interested in are constructed
out of powers of the left-handed current $L_\mu $, more precisely, from
powers of the commutators $f_{\mu \nu }\equiv [L_\mu ,L_\nu ]$. For example,
the lowest order contributions may have the form
\begin{equation}
\label{Traces}%
\mathop{\rm Tr}
f_{\mu \nu }f^{\mu \nu },\quad
\mathop{\rm Tr}
f_\mu ^{\ \nu }f_\nu ^{\ \lambda }f_\lambda ^{\ \mu },\quad
\mathop{\rm Tr}
\left( f_{\mu \nu }f^{\mu \nu }\right) ^2,\quad
\mathop{\rm Tr}
f_\mu ^{\ \nu }f_\nu ^{\ \lambda }f_\lambda ^{\ \sigma }f_\sigma ^{\ \mu
},\quad \quad {\it etc...}
\end{equation}
The choice of such combinations is motivated by the possibility that they
could be induced by hidden gauge symmetry (HGS) terms \cite
{Bando85,Marleau89-10,Marleau93-11}. They are also automatically chirally
invariant. Chiral symmetry breaking must be added independently, usually by
adding a pion mass term, which means that one has in principle more control
on the symmetry breaking mechanism. The Skyrme term itself emerges from the
gauge field kinetic term in the limit of large gauge vector mass in this
formulation.

Using the hedgehog ansatz for the static solution, i.e. $U=\exp [i{\bf \tau
\cdot }\widehat{r}F(r)]$ where $F(r)$ is the chiral angle, a number of
properties of the traces were found (see ref. \cite{Marleau90-13}):

\begin{description}
\item[(i)]  All traces are polynomials of the following combinations, $%
a\equiv r^{-2}\sin ^2F$ and $b\equiv F^{\prime 2}$. They can be written as
\begin{equation}
\label{fn}%
\mathop{\rm Tr}
(f_{\mu \nu })^n={\rm \ }\sum_{m=0}^M\kappa _{n,m}a^{n-m}(b-a)^m
\end{equation}
where $(f_{\mu \nu })^n$ represents a linear combination of any Lorentz
invariant product of $n$ $f_{\mu \nu }$'s and $M=\left[ \frac n2\right] $ is
the integer part of $\ \frac n2.$

\item[(ii)]  The constants $\kappa _{n,m}$ are such that the ration of $%
\kappa _{n,0}$ over $\kappa _{n,1}$ is found to be $\frac 3n$. For example,%
$$
\mathop{\rm Tr}
f_{\mu \nu }f^{\mu \nu }=16a[a+2b],\quad
\mathop{\rm Tr}
f_{\mu \nu }f^{\nu \lambda }f_\lambda ^{\ \ \mu }=96a^2b,...\quad
$$
where the first expression corresponds to the Skyrme term. Note that $%
\mathop{\rm Tr}
L_\mu L^\mu $ in (\ref{LSkyrme}) also gives a similar result, i.e. $%
\mathop{\rm Tr}
L_\mu L^\mu =2(2a+b)$. Furthermore, one can construct a special class \cite
{Marleau89-10} of such combinations which is at most linear in $b$ (or of
degree two in derivatives of $F$). These Lagrangians give a very simple form
for the hedgehog solution
\begin{equation}
\label{fn2}%
\mathop{\rm Tr}
(f_{\mu \nu })^n=\kappa _{n,0}{\rm \ }a^{n-1}[3a+n(b-a)]
\end{equation}
and lead to a chiral angle equation which is tractable since it is of degree
two. It also turns out that for this class of Lagrangians ${\rm \kappa }%
_{n,0}=0$ for $n$ odd $\geq 5$.
\end{description}

The energy of the static solution (or mass of the soliton) is then given by
the general expression:
\begin{equation}
\label{Mstatic}{\cal M}_S\equiv -\int d^3r\ {\cal L}=4\pi \int
r^2dr\sum_{n=1}^\infty \sum_{m=0}^M\kappa _{n,m}a^{n-m}(b-a)^m.
\end{equation}
The special class of combinations which is at most linear in $b$ gives an
even simpler form since $\kappa _{n,m}=0$ for $m>1$:
\begin{equation}
\label{Xi}{\cal M}_S=4\pi \int r^2dr\left[ 3\chi (a)+(b-a)\chi ^{\prime
}(a)\right]
\end{equation}
where $\chi (x)=\sum_{n=1}^\infty c_nx^n$ and $\chi ^{\prime }(x)=d\chi
(x)/dx$. Note that for ${\cal M}_S$ to be positive, it is sufficient (but
necessary) to impose the conditions $c_n\geq 0$ and $(b-a)$ $\geq 0$.
Minimizing the static energy with respect to the chiral angle leads to
differential equation:
\begin{equation}
\label{chiral}\chi ^{\prime }(a)(F^{\prime \prime }+2\frac{F^{\prime }}r-2%
\frac{\sin F\cos F}{r^2})+a\chi ^{\prime \prime }(a)(F^{\prime 2}\cot F-2%
\frac{F^{\prime }}r+\frac{\sin F\cos F}{r^2})=0
\end{equation}
where again $a\equiv r^{-2}\sin ^2F$.

The existence of solutions requires that the series $\chi (a)$ converges for
all values of $r$, leading to one can call {\bf summable} Skyrmions. So it
is closely related to the scaling properties of the Lagrangian. Order by
order the terms scales according to
\begin{equation}
\label{Mtot}{\cal M}_S\equiv -\int d^3r\ {\cal L}=\alpha _2R+\alpha
_4R^{-1}+\alpha _6R^{-3}+\cdots +\alpha _nR^{3-n}+\cdots
\end{equation}
when $r$ is scaled according to $r\rightarrow rR^{-1}$ and 2$n$ is the
number of derivatives. Obviously, the first term prevent the soliton from
exploding. The exact behavior as $R\rightarrow 0$ and the existence of a
global minimum (and of a Skyrmions) depend on the details of the series $%
\chi (x).$ The existence of Skyrmions has been demonstrated for a number of
models for which it is also possible to compute baryon static properties
\cite{Marleau90-13,Riska93,Jackson91} using usual methods. In some cases,
the solutions have shown an interesting two-phase structure. This last
behavior is related to the convergence of $\chi (x)$.

The Lagrangians in (\ref{fn2}) must be constructed by hand, order by order.
But the contribution to the static energy obeys a recursion relation. If $%
{\cal M}_n=a^{n-1}[3a+n(b-a)]$ is the static energy coming from order 2$n$,
then the $n^{th}$ contribution could be obtain from%
$$
{\cal M}_n=-{\cal M}_{n-2}{\cal M}_2+{\cal M}_{n-4}{\cal M}_4-\frac 13{\cal M%
}_{n-6}{\cal M}_6.
$$
for even $n\geq 8$.

Finally, we mention that similar results hold for constructions of the form $%
\mathop{\rm Tr}
(L_\mu )^{2n}$ but in this case $M=n$ in the expression in (\ref{fn2}) and
in general, ${\rm \kappa }_{n,0}\neq 0$ for $n$ odd $\geq 5$. A
corresponding recursion relation, which applies in this case to all $n>4,$
can also be written.

\subsection{New classes}

In light of those results we propose two other classes of all-orders
Skyrmions. First, let us rewrite equation (\ref{fn}),
\begin{equation}
\label{fnb}%
\mathop{\rm Tr}
(f_{\mu \nu })^n=\kappa _{n,0}[a^n+(2a^n+na^{n-1}(b-a)+\cdots +\frac{\kappa
_{n,M}}{\kappa _{n,0}}a^M(b-a)^M)].
\end{equation}
where $M=[\frac n2]$. As mentioned above, it is possible to combine several
traces of order $n$ and eliminate higher orders in $b$ and obtain a class of
summable Skyrmions. But the last expression also suggest another interesting
case where the traces combine in a way that the term in parenthesis
corresponds to the binomial expansion. It simplifies as follows:
\begin{equation}
\label{fnb2}%
\mathop{\rm Tr}
(f_{\mu \nu })^n=\kappa _{n,0}[a^n+2a^M(a+(b-a))^M)]=\kappa
_{n,0}[a^{2M}+2(ab)^M]
\end{equation}
for $n$ even and $M=\frac n2$.

This {\bf second class} of all-orders Skyrmions is then characterized by the
static energy
\begin{equation}
\label{Xi2}{\cal M}_S=4\pi \int r^2dr\left[ \chi (a)+2\chi (\sqrt{ab}%
)\right]
\end{equation}
where $\chi (x)=\sum_{n=1}^\infty c_nx^n$ with $c_n=0$ for $n$ odd $>3$.
Here for ${\cal M}_S$ to be positive definite, it is sufficient that the
model obeys $c_n\geq 0$ . The chiral angle is then described by:
\begin{equation}
\label{chiral2}2\chi ^{\prime }(a)\cos F-\chi ^{\prime }(\sqrt{ab})-\chi
^{\prime \prime }(\sqrt{ab})(F^{\prime 2}\cos F+F^{\prime \prime }\sin
F-F^{\prime }\frac{\sin F}r)=0
\end{equation}
where again $a\equiv r^{-2}\sin ^2F$ and $b\equiv F^{\prime 2}$. This second
class of Lagrangians is generated order by order by the systematic procedure
(see ref. \cite{Marleau90-13} for details):
\begin{equation}
\mathop{\rm Tr}
(f_{\mu \nu })^n=%
\mathop{\rm Tr}
[\{f_\mu {}^\nu ,f_\lambda {}^\rho \}\{f_\nu {}^\lambda ,f_\sigma {}^\omega
\}\{f_\rho {}^\sigma ,f_\xi {}^\eta \}\{f_\omega {}^\xi ,f_\eta {}^\beta
\}\cdots ].
\end{equation}

The {\bf third class} of all-orders Skyrmions comes on the other hand from
Lagrangians build in terms of even powers of $L_\mu $'s instead of powers of
$f_{\mu \nu }$'s. In this case, the traces contains terms up to $O(b^n)$ and
can be written
\begin{equation}
\label{Lnb}%
\mathop{\rm Tr}
(L_\mu )^{2n}=\kappa _{n,0}[2a^n+(a^n+na^{n-1}(b-a)+\cdots +\frac{\kappa
_{n,n}}{\kappa _{n,0}}a^n(b-a)^n)].
\end{equation}
Following equation (\ref{fnb2}), we choose the combination for which the
term in parenthesis correspond to the binomial expansion of $(a+(b-a))^n$
and find
\begin{equation}
\label{fnb3}%
\mathop{\rm Tr}
(L_\mu )^{2n}=\kappa _{n,0}(2a^n+b^n).
\end{equation}
The explicit structure of traces leading to the above result is the a simple
product of the anticommutator defined by $d_{\mu \nu }\equiv \{L_\mu ,L_\nu
\}$, i.e.%
$$
\mathop{\rm Tr}
d_\mu ^{\ \nu }d_\nu ^{\ \lambda }d_\lambda ^{\ \sigma }d_\sigma ^{\ \alpha
}\cdots d_\rho ^{\ \mu }=\kappa _{n,0}(2a^n+b^n).
$$

The static energy takes a simple form
\begin{equation}
\label{Xi3}{\cal M}_S=4\pi \int r^2dr\left[ 2\chi (a)+\chi (b)\right]
\end{equation}
where $\chi (x)=\sum_{n=1}^\infty c_nx^n$. For ${\cal M}_S$ to be positive
definite, it is sufficient that $c_n\geq 0$ . The chiral angle must now
obey:
\begin{equation}
\label{chiral3}2\chi ^{\prime }(a)\sin F\cos F-\chi ^{\prime
}(b)(2rF^{\prime }+r^2F^{\prime \prime })-2\chi ^{\prime \prime
}(b)r^2F^{\prime 2}F^{\prime \prime }=0
\end{equation}
This last expression takes a very simple form if we define $\zeta (x)\equiv
\chi (x^2)=\sum_{n=1}^\infty c_nx^{2n}$ since the arguments of $\chi $ are
squared objects (either $a$ or $b$). The equation becomes
\begin{equation}
\label{chiral3a}2\zeta ^{\prime }(\frac{\sin F}r)r\cos F-2\zeta ^{\prime
}(F^{\prime })-\zeta ^{\prime \prime }(F^{\prime })r^2F^{\prime \prime }=0.
\end{equation}
The results in (\ref{chiral}), (\ref{chiral2}) and (\ref{chiral3a}) can be
considered as special cases. In general, the equation for the chiral angle
should depends on a more complicated functional ,{\bf \ }$\chi (a,b),$ to
take into account the general expression in (\ref{fnb}).

\section{Toy models with analytical solutions}

Instead of trying to find solutions for each class of model, we adopt a
different point of view here. We look at special solutions and try to find
what model can accommodate it at the expense of physical meaning. The
interest of the procedure will become more clear below. For this purpose, we
will assume that the function $\chi (x)$ is no longer constrained to have a
power series representation so are not directly related to the
aforementioned classes of Lagrangians. Looking at the chiral angle equations
described above its becomes apparent that a solution which obeys
\begin{equation}
\label{solutionF}F^{\prime }=-\alpha \frac{\sin F}r,
\end{equation}
simplifies the chiral angle equation for all class of models. Since we are
interested in the N=1 soliton with $F(0)=\pi $ and $F(\infty )=0$, the
constant $\alpha $ must be positive. The solution for $F$ is simply:

\begin{equation}
\label{arctan}F(r)=2\arctan (a_1r^{-\alpha })
\end{equation}
Since the static energy must be finite, $F$ is required to decrease fast
enough as $r\rightarrow \infty .$ Clearly this imposes an extra condition on
the range of possible values of $\alpha .$

Imposing this solution for the chiral angle equations in (\ref{chiral}), (%
\ref{chiral2}) and (\ref{chiral3a}) translate into differential equations
for $\chi (x).$ In the first case, equation (\ref{chiral}) becomes%
$$
(\alpha ^2-1)x\chi ^{\prime \prime }(x)=-(3-2\alpha ^2)\chi ^{\prime }(x).
$$
which can be solved for $\chi (x):$%
\begin{equation}
\label{xi1}\chi (x)=C_1+\frac{C_2}\beta x^\beta
\end{equation}
with $\beta =\frac{3\alpha ^2-4}{\alpha ^2-1}.$ Physically plausible
solution requires that $C_1=0$ because of the finiteness of the static
energy. The remaining model corresponds to a Lagrangian with 2$\beta $ in
derivatives of the pion field. In order to make any sense $2\beta $ must be
an integer, otherwise one is forced to consider fractional power of a
derivative. From scaling argument, it is easy to see that for $\beta <\frac 3%
{2,}$ the ''size'' of the solution collapse to zero, for $\beta >\frac 32$,
energy favors infinite size configuration and the solution eventually
disperse. Therefore none of those solution are stable solitons. However for $%
\beta =\frac 32$, the solution is scale invariant. Among the values that $%
\beta $ may take, three of them seem of particular interest:

\begin{description}
\item[(i)]  $\alpha =\sqrt{\frac 32}$: For this value of $\alpha ,$ $\beta
=1 $ and the model correspond to the nonlinear $\sigma $-model. The solution
is therefore not stable under scale transformations but at least one can
attach a physical meaning to the model.

\item[(ii)]  $\alpha =1$: This solution correspond to the limit of $\beta
\rightarrow \infty ,$ i.e. the only contribution comes from a term with an
infinite number of derivatives. The solution is not a stable Skyrmion.

\item[(iii)]  $\alpha =\sqrt{\frac 53}$: Since $\beta =\frac 32$, the static
energy given by this model
\begin{equation}
\label{Xi3/2}{\cal M}_S=4\pi C_2\int r^2dr\left( \frac{\sin ^2F}{r^2}\right)
^{\frac 12}\left[ \frac{\sin ^2F}{r^2}+F^{\prime 2}\right]
\end{equation}
is invariant under scaling. Using the solution, we can rewrite the static
energy
\begin{equation}
\label{Ms3/2}{\cal M}_S=-4\pi C_2\cdot \frac 8{\sqrt{15}}\cdot \int_0^\infty
drF^{\prime }\sin ^2F=\frac{16\pi ^2C_2}{\sqrt{15}}\cdot N
\end{equation}
and we see that result is proportional to the topological number, $N=1$
here. The connection between this model and any physically relevant
Lagrangian is not obvious, but the mere fact that the static energy is
proportional to the topological (or baryon) number is in itself rather
interesting.
\end{description}

The second class of summable Skyrmions, described by equation (\ref{chiral2}%
), requires some care. In order to simplify the differential equation, we
will assume that $\chi $ takes a form similar to (\ref{xi1}). For this
choice to be consistent one must then look at the negative root of $\sqrt{ab}
$, i.e.%
$$
\sqrt{ab}=-\alpha \frac{\sin ^2F}{r^2}
$$
in which case $\beta =\frac 12$, and $\alpha $ is found to be $4$ for the $%
N=1$ soliton solution. If we were to associate this model to a Lagrangian we
would have to construct one with a single, i.e. 2$\beta =1,$ derivative of
the field. In any case, there is no stable Skyrmion here but the energy of
this unstable solution takes the simple form
\begin{equation}
\label{Ms1}{\cal M}_S=24\pi C_2\int drr\sin F=\frac{48\pi }{\sqrt{a_1}}%
C_2\cdot \frac \pi {4\sqrt{2}}
\end{equation}

Finally, the equation for the third class of summable Skyrmions is given by (%
\ref{chiral3}) and for $\alpha =1$, it becomes
\begin{equation}
2x\chi ^{\prime \prime }(x)=\chi ^{\prime }(x).
\end{equation}
Solving for $\chi (x)$, we get the form (\ref{xi1}) with $\beta =\frac 32$
(and $C_1=0$). The function $\chi (x)$ is identical to the one found in the
first class with the choice $\alpha =\sqrt{\frac 53}$. This means that the
static energy is again invariant under scaling. Although the static energy
can be recast in a form similar to (\ref{Ms3/2}), i.e. proportional to the
topological number, the chiral angle is clearly not the same:
\begin{equation}
\label{arctan3}F(r)=2\arctan (a_1r^{-1})
\end{equation}
and the two models are completely different.

\section{Discussion and conclusions}

We have introduce two new classes of summable Skyrmions and have written the
Lagrangians which induce them. The basic elements for solving the chiral
angle equation are in place and once the solution is found (by numerical
analysis usually) the physical quantities can computed in a straightforward
procedure similar to the first class of models. There remains however a
noumber of open question. One must verify that solution can be reach for any
plausible model and provide examples where these new summable Skyrmions
indeed exists. The second point that would be interesting to analyse is the
possibility of a phase structure. In the first class of model, the phase
structure is related to the convergence of the series for $\chi (a)$. In the
second and third classes $\chi (\sqrt{ab})$ and $\chi (ab),$ respectively,
are also required to converge otherwise the solution has infinite energy.
The extra requirement may have some intriguing effects on the solutions. It
may well be that it will introduce a solution which experience three phases,
one dominated by small $a$ and $b$ a second one regulated by the convergence
of $\chi (a)$ and a new phase regulated by the convergence of $\chi (\sqrt{ab%
})$ or $\chi (ab)$. One can also imagine another alternative: a bifurcation
between two solutions each having a two-phase structure, the first solution
would correspond to $\chi (a)$ and the second, to $\chi (\sqrt{ab})$ or $%
\chi (ab)$. In principle the two solutions could even be degenerate. Since
there is still a large degree of freedom in the construction of physically
viable models none of these effects are excluded a priori. Finally, the
question of quantum stability must also be addressed for these new classes
of models.

Finally, we have also found analytical solution to toy models in this work.
They are motivated by the relations that were obtained for the class of
summable Skyrmions but no longer carries any relations with the power series
representation or with the Lagrangian constructed in section II, and so,
their physical interpretation is unclear. However the main conclusion that
comes out of this exercise is that the stable annalytical solutions
(solitons) that were found are scale-invariant solutions and their static
energy proportional to the topological number which is a feature also
observed in the sine-Gordon model in (1+1) space-time dimensions.

\section{Acknowledgement}

The author would like to thank T.N. Truong for the hospitality of the Centre
de Physique Th\'eorique where part of this work was done. This research was
supported by the Natural Science and Engineering Research Council of Canada
and by the Fonds pour la Formation de Chercheurs et l'Aide \`a la Recherche
du Qu\'ebec.

\end{document}